\documentclass[12pt,a4paper,oneside]{book}

\usepackage{lipsum}
\usepackage{mathtools}
\usepackage[round]{natbib}
\usepackage{graphicx}
\usepackage{algorithm}
\usepackage{algorithmic}
\usepackage{markdown}
\usepackage[none]{hyphenat}
\usepackage{booktabs}
\usepackage{gensymb}

\raggedright

\numberwithin{section}{chapter}
\numberwithin{table}{chapter}
\numberwithin{figure}{chapter}
\numberwithin{equation}{chapter}

\graphicspath{ {images/} }

\parskip = \baselineskip
\title{Predicting Malaria Incidence using Neural Networks and Disaggregation Regression}
\author{Jack A. Hall}

\begin{document}

\begin{titlepage} 
	\newcommand{\HRule}{\rule{\linewidth}{0.5mm}} 
	
	\center 
	
	
	\textsc{\LARGE University of Leicester}\\[0.5cm] 
	\textsc{\Large Department of Health Sciences}\\[1.5cm]
	
	\textsc{\Large MSc Medical Statistics}\\[0.5cm] 
	
	
	
	\HRule\\[1.5cm]
	
	{\huge\bfseries Predicting Malaria Incidence Using
	Artifical Neural Networks and Disaggregation Regression} 
	
	\HRule\\[1.5cm]
	
	
	\begin{minipage}{0.4\textwidth}
		\begin{flushleft}
			\large
			\textit{Author}\\
			Jack A. \textsc{Hall}\\
			0000-0002-3016-7624
		\end{flushleft}
	\end{minipage}
	~
	\begin{minipage}{0.4\textwidth}
		\begin{flushright}
			\large
			\textit{Supervisor}\\
			Dr. Tim C.D. \textsc{Lucas}\\
			0000-0003-4694-8107
		\end{flushright}
	\end{minipage}
	
	
	
	\vfill\vfill\vfill 
	
	{\large\today} 
	
	
	 
	
	\vfill 
	
\end{titlepage}

\frontmatter

\section*{Abstract}
\label{ch:abstract}

\paragraph*{Background}

Disaggregation modelling is a method of predicting disease risk at high resolution using aggregated response data. High resolution disease mapping is an important public health tool to aid the optimisation of resources, and is commonly used in assisting responses to diseases such as malaria. Current
disaggregation regression methods are slow, inflexible, and do not easily allow non-linear terms. 

\paragraph*{Aim}

Neural networks may offer a solution to the limitations of current disaggregation methods. This project aimed to design a neural network which mimics the behaviour of disaggregation, then benchmark it against current methods for accuracy, flexibility and speed.

\paragraph*{Methods}

Cross-validation and nested cross-validation tested neural networks against traditional disaggregation for accuracy and execution speed was measured.

\paragraph*{Results}

Neural networks did not improve on the accuracy of current disaggregation methods, although did see an improvement in execution time. The neural network models are more flexible and offer potential for further improvements on all metrics. The R package 'Kedis' (Keras-Disaggregation) is introduced as a user-friendly method of implementing neural network disaggregation models.

\section*{Acknowledgements}
\label{ch:acknowledgements}
I would like to thank the person that discovered coffee. Without you, this work would not have been possible. I would also like to thank Megan for obtaining the equipment required to inject it directly into my veins.

\tableofcontents
\listoffigures
\listoftables

\mainmatter

\chapter{Introduction}
\label{ch:introduction}
With 350–500 million cases leading to up to one million deaths per year, malaria remains one of the deadliest infectious diseases in the world. It is widespread in all tropical and subtropical areas \citep{ponts2013}. Funds from international sources committed towards malaria control in 2012 was \$1.84 billion \citep{alonso2013}, although it is still not possible to provide all interventions to all areas at risk \citep{drake2017}. It is therefore necessary to develop tools to help optimise allocation of resources to areas at the greatest risk. High resolution maps of disease risk are an important public health tool, facilitating efficient allocation of limited resources and precision targeting of interventions \citep{arambepola2021}. Often, disease cases are reported as aggregated data across regions, such as states or districts, and this aggregation is problematic for fine scale predictions due to the ecological fallacy, where relationships between covariates may not be the same at all scales \citep{arambepola2021}. 

Disaggregation regression is a method of making predictions at a high resolution given a low resolution response data. In terms of disease mapping, it can take aggregated disease cases and produce a map of disease risk at a fine spatial scale, using covariates such as temperature or humidity at that same high scale. It was first introduced as a method of species distribution modelling in ecology \citep{keil2013} and is now commonly used in disease mapping, particularly with malaria \citep{weiss2019, battle2019}. 

\section{Disaggregation Regression}
Disaggregation regression is an extension of generalised linear models (GLMs), where the outputs are ‘aggregated’ over regions to form one prediction for each region. A traditional GLM is typically expressed as:

\begin{equation} \label{eq:glm}
    g\left(\mu\right)=\eta=\beta_0+\beta_1x_1+\cdots+\beta_nx_n
\end{equation}

For malaria modelling, we often have total cases reported across regions in a pre-specified time period, such as one year, so a Poisson distribution can be used to model the distribution. Poisson regression typically uses a log link function, so Equation~\ref{eq:glm} can be rewritten as:

\begin{equation} \label{eq:pois_glm}
    \log(\mu)=\eta=\beta_0+\beta_1 x_1+\cdots+\beta_n x_n
\end{equation}

Where \(\eta\) is the linear predictor, \(\mu\) is the estimate for the number of cases of malaria per person per unit time, \(x_n\) are the values for the covariates, and \(\beta_n\) represent the coefficients for each parameter. If the population \(p\) is known, then the rate (total cases per unit time) \(y=\mu p\) can be derived:

\begin{equation} \label{eq:pois_glm_offset}
\begin{split}
        \log(y)&=\eta+\log(p)\\&=\beta_0+\beta_1x_1+\cdots+\beta_nx_n+\log(p)
\end{split}
\end{equation}

\begin{equation} \label{eq:pois_glm_offset_exp}
    y=\mu p=e^{\beta_0+\beta_1x_1+\cdots+\beta_nx_n+\log(p)}
\end{equation}

Where \(\log(p)\) is referred to as the ‘offset,’ or the part of the linear predictor that is known.

Disaggregation regression differs from traditional regression in that the cases are aggregated over a large area. Covariates such as temperature or humidity are truly continuous but are represented by computers as a grid of pixels, akin to pixels on a television. Thus, for a region with \(n\) pixels, the total cases per unit time is the sum of all the cases in each pixel:

\begin{equation} \label{eq:agg_pois_glm}
\begin{split}
 y&=\sum_{i=1}^{n}\left.\left[e^{\left(\beta_0+\beta_1{x_1}_i+\cdots+\beta_n{x_n}_i+\log{\left(p_i\right)}\right)}\right]\right.\\&=\sum_{i=1}^{n}\left.\left[e^{\left(\beta_0+\beta_1{x_1}_i+\cdots+\beta_n{x_n}_i\right)}p_i\right]\right.
\end{split}
\end{equation}

Equation~\ref{eq:agg_pois_glm} describes a case where the relationship between covariates \(x\) and the response \(y\) is the same for every pixel, with no correlation between adjacent pixels that is not explained by the covariates. In reality, this is an important effect that should be modelled, so a Gaussian process may be used. Gaussian processes are a method of estimating a continuous field from a discrete set of data points, and can be added to a linear predictor of a disaggregation model to adjust for spatial autocorrelation. A Gaussian process is equivalent to a single-layer fully-connected neural network with an i.i.d. prior over its parameters, in the limit of infinite network width \citep{lee2017}. Approximations of a Gaussian process with a single layer of a finite (but large) width may be useful in disaggregation modelling with a neural network.

\section{Disaggregation 'R' Package}
Disaggregation regression can be implemented in R using the ‘disaggregation’ package \citep{disaggregation}. It uses the INLA, or Integrated Nested Laplace Approximation, method \citep{INLA1, INLA2, INLA3, INLA4, INLA5, INLA6, INLA7, INLA8, INLA9} to approximate the Gaussian Process and TMB, or Template Model Builder, which provides efficient functionality for random effects models and differentiation in C++ \citep{TMB1, TMB2}. 

The disaggregation package makes it easy for the user to run disaggregation models at the expense of some flexibility \citep{disaggregation}. Although ‘disaggregation’ offers vast speed improvements over previous methods such as MCMC, it is still slow and cumbersome. Disaggregation is limited in its ability to model non-linear relationships and does not offer a user friendly option to modify the models beyond those directly specified in the package. It is necessary to manually specify more complex models.

The disaggregation package fits a disaggregation regression using spatial data provided into R via the \verb|raster| package, and creates predictions with uncertainty of the class \verb|RasterLayer|. The Gaussian process is estimated by the INLA package, which builds a mesh from a set of triangles from the datapoints then interpolates between. Hyperparameters determine the size of the triangles; smaller triangles offer a better accuracy but are much slower to fit than larger triangles. An IID random effect can be included as noise and added to the linear predictor at each pixel, and is intended to improve the model fit by adjusting for overdispersion. The IID variables have a mean of 0 and the standard deviation is updated from a (normally vague) prior during model fitting.

\section{Artificial Neural Networks}
\begin{figure} 
    \centering
    \includegraphics[width=\textwidth]{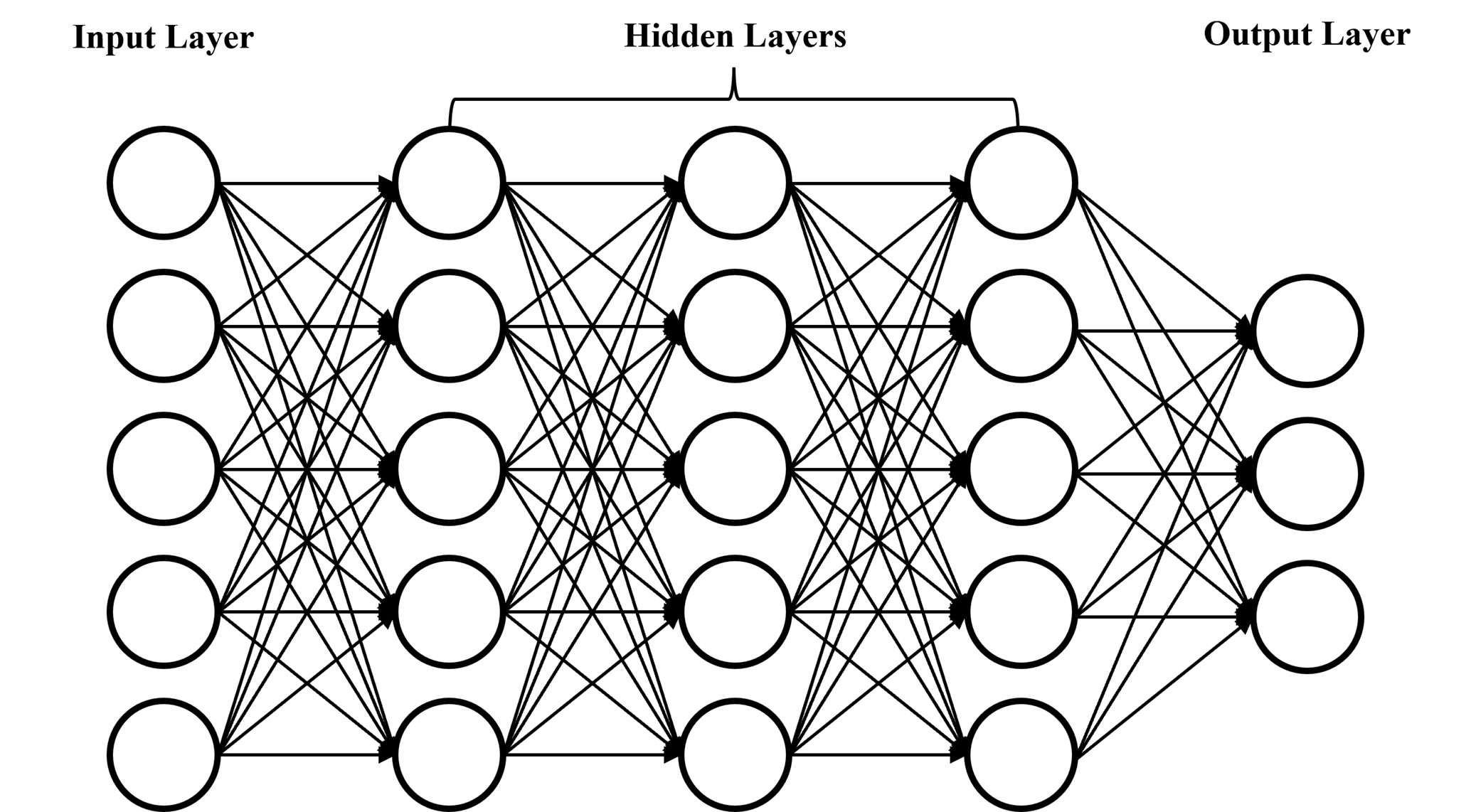}
    \caption{An example artificial neural network}
    \label{fig:ann}
\end{figure}

Implementing disaggregation regression using a neural network is one possibility for fixing the lack of non-linear functions, slow speed and reduced flexibility of the ‘disaggregation’ package. Artificial neural networks (ANNs) are a type of supervised machine learning algorithm inspired by neurons in the human brain. Nodes are interconnected by pathways, and those pathways ‘learn’ weights based on the information it is given through back-propagation using an algorithm defined by the optimiser. Typically, these nodes are arranged in multiple layers and each layer is fully interconnected with the preceding and succeeding layers (Figure~\ref{fig:ann}). The output of each node can be thought of as a linear combination of all the inputs (plus a bias, which is equivalent to an intercept in a linear regression), all multiplied by an activation function. The stacking of these linear regression models is what gives ANNs enormous power to solve complex classification or regression problems.

\subsection{Activation Function}
The output of each node in an ANN is subject to a transformation to add non-linearity to the model. Without activation functions, the output of a neural network can only be a linear combination of the input nodes, as the sum of multiple linear functions is also a linear function. Common activation function are Rectified Linear Unit (ReLU), sigmoid (the inverse logit function) or Hyperbolic Tangent (tanh). The Rectified Linear Unit is a simple transformation where the output is zero if the input is negative, and the equal to the input otherwise. It can be expressed mathematically as:

\begin{equation} \label{eq:relu}
    f(x)=max(0,x)
\end{equation}

The sigmoid activation function is a transform that bounds the output between 0 and 1, and can be useful in scenarios where odds are a desired score (it is the inverse of the logit function). It is represented mathematically as:

\begin{equation} \label{eq:sigmoid}
    f(x)=\frac{1}{1+e^{-x}}
\end{equation}

The hyperbolic tangent (Tanh) activation function is similar to sigmoid, although it is stretched vertically so is bounded between -1 and 1, and pushes input values closer to the extremes faster than sigmoid. It is defined as:

\begin{equation}
f(x)=\frac{e^x-e^{-x}}{e^x+e^{-x}}
\end{equation}

Although the myriad of activation functions all apply different transformations to the output of a node, the general purpose is always to add non-linearity to a model or, in the case of the final output node, transform the output to a range that makes sense (similar to the purpose of a link function in GLMs). The rectified linear activation function, or ReLU activation function, is perhaps the most common function used for hidden layers \citep{brownlee_2021}.

\subsection{Loss Function}

Loss is a key part of a machine learning model, as it describes how well the model fits and influences the optimisation process. The loss function is a hyperparameter to be decided by the model designer and must consider the activation functions and purpose of the model. Loss functions are related to maximum likelihood estimation in traditional statistical modelling and the process of iterating to find an optimal parameter set given a set of data is similar in both cases. In many cases, loss functions are directly related to the likelihood of the relevant distribution, so minimising the loss is mathematically equivalent to finding the maximum likelihood.

Where the response variable is a discrete count, as in the case of malaria case modelling, a Poisson distribution can be used so the Poisson loss is used as the loss function. Minimising the Poisson loss is equivalent to maximising the likelihood of a Poisson regression. Poisson loss is defined as:

\begin{equation}
    L\left(y,\hat{y}\right)=\frac{1}{n}\sum_{i=0}^{n}\left[\widehat{y_i}-y_i\log{\left(\widehat{y_i}\right)}\right]
\end{equation}

Other loss functions could be used including Root Mean Squared Error or Mean Absolute Error, and the choice depends on the particular model.

\subsection{Optimiser}

The optimiser describes the algorithm which is used to adjust the weights of a model in order to minimise the loss. A simple example of an optimiser is Gradient Descent, which uses the first derivative of the loss function to determine which direction to adjust each weight in order to minimise the loss. This is the method most often used in linear regressions, and has the advantage that it is easy to compute, implement and understand. The disadvantages of this method are that it can be slow, especially in large datasets, and may trap local minima. Alternative optimisers exist which aim to resolve these issues, each with advantages and disadvantages, and it is the job of the model designer to decide on an optimiser that works well for their specific problem. One alternative optimiser is the Adam optimiser \citep{kingma2014}, which uses the concept of 'momentum' (or the second derivative of the loss function) and adjusting the learning rate by a running average of the recent weights to improve the speed of optimisation and to avoid getting trapped in local minima.

\subsection{Neural Network Implementations}

Although it is possible to build neural networks from scratch, there are 'prepackaged' implementations that aim to accelerate the development process by removing some trivial steps and provides a user-friendly API. Some common implementations are TensorFlow \citep{abadi2016tensorflow}, Keras \citep{chollet2015keras}, and PyTorch \citep{paszke2019pytorch}. Keras and TensorFlow have packages in R \citep{keras, tensorflow}, whereas PyTorch only has interfaces for Python, C++ and Julia. Keras is built upon TensorFlow, and designed as a more user-friendly version of TensorFlow to speed up the development cycle, so for that reason it has been chosen as the implementation for this project.

\section{Spatial Data}
Spatial phenomena can generally be thought of as either discrete objects with clear boundaries or as a continuous phenomenon that can be observed everywhere but does not have natural boundaries \citep{hijmans2021}. In the context of disaggregation mapping, regions by which cases are aggregated can be thought of as discrete regions, and covariates such as temperature or elevation are continuous. Discrete and continuous data are represented as vectors and rasters, respectively.

\subsection{Vectors}
The main vector data types are points, lines and polygons. In all cases, the geometry of these data structures consists of sets of coordinate pairs (x, y) \citep{hijmans2021}. In disaggregation, we are usually only interested in the polygon data, as this is what is used to represent regions over which cases are aggregated. A vector data file can be thought of as a data frame, with a column of geometry data (which completely describes the boundaries of the area), and additional columns with additional data for each region, such as region name, code, or cases of malaria reported.

\subsection{Rasters}
Raster data represents continuous spatial phenomenon such as temperature or elevation. Rasters divide the region of interest into grids with rectangles (called cells or pixels) that contain the variable of interest. The resolution of a raster represents how finely the true continuous picture is divided into squares, with smaller squares being a higher resolution. In disaggregation regression, the covariates of the model are rasters, and the prediction will be a raster of the same resolution as the covariates.

\subsection{R Packages for Handling Spatial Data}

Base R does not offer methods for handling spatial data, so external packages are necessary. The disaggregation package uses the \verb|raster| package \citep{raster} and the \verb|sp| package \citep{sp1, sp2}, which have been superseded by \verb|terra| \citep{terra} and \verb|sf| \citep{sf}. All packages handle data in similar ways, but the newer packages are generally faster and offer more functionality.

\section{The Madagascar Dataset}

The dataset used in this  project is comprised of a set of covariate rasters, a raster of population at the same resolution as the covarites, and a shapefile with administrative borders and counts of cases reported within those regions over a fixed period of time. The response data was provided by the National Malaria Control Program in Madagascar (NMCP) who gave the number of patients who tested positive for Malaris with Rapid Diagnostic Tests (RDTs), irrespective of species. The counts were pre-processed to take into consideration treatment seeking behaviour prior to receiving the dataset. Covariates were obtained from NASA's MODIS project \citep{modis}, and population data was obtained from \citet{linard2012}.

\begin{figure}
    \centering
    \includegraphics[width=\textwidth]{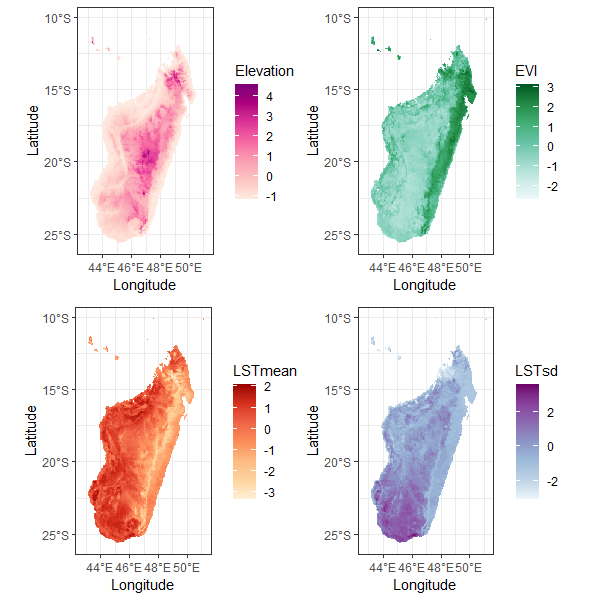}
    \caption{Maps of the (normalised) covariate rasters used in this project}
    \label{fig:covariates}
\end{figure}

\begin{figure}
    \centering
    \includegraphics[width=\textwidth]{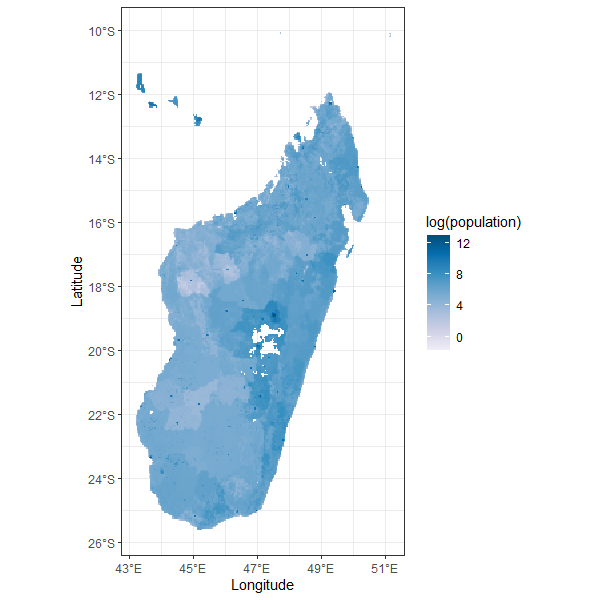}
    \caption{Map of the log population raster of Madagascar}
    \label{fig:population}
\end{figure}

\begin{figure}
    \centering
    \includegraphics[width=\textwidth]{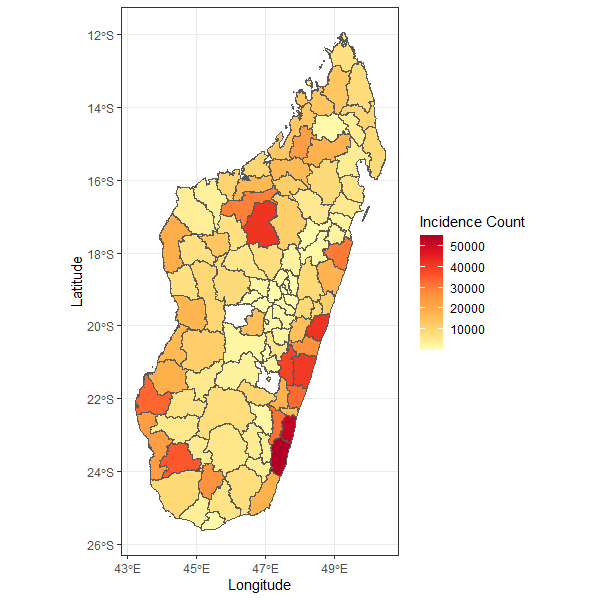}
    \caption{Map of the incidence of malaria reported by administrative region in Madagascar}
    \label{fig:counts}
\end{figure}

The covariates used in this project were Elevation, Enhanced Vegetation Index (EVI), Land Surface Temperature (LST) mean, and Land Surface Temperature (LST) standard deviation, all of which have shown to be useful covariates in malaria modelling \citep{lucas2022}. Figure~\ref{fig:covariates} displays the four covariates used in the model, after scaling and centering. The resolution of the rasters is 24 pixels per degree, or 2.5 square nautical miles (approximately 4km\(\times\)4km). The population raster is displayed as Figure~\ref{fig:population}, although it has been transformed to a log scale for better visualisation. It is at the same resolution as the covariates. Regions of high population are centred around cities, with the capital city Antananarivo at approximately 19\degree S, 47\degree E having the highest population density. White areas show missing data. Figure~\ref{fig:counts} shows the administrative borders of Madagascar and the reported cases within those borders. Darker red areas reported a higher number of cases, although this may be explained by a larger population (the offset term in the Poisson regression).

\section{Outline of Project Aims}
This primary aim of this project is to design, build, and test a neural network implementation of disaggregation regression. This novel method will then be tested against the existing disaggregation implementation to investigate whether it has improved accuracy, greater flexibility and faster computation time. Finally, methods of dealing with uncertainty in the predictions will be investigated.

\chapter{Methods}
\label{ch:methods}
In this section, the architecture of a Kedis model is examined in detail. The methods used to test Kedis against disaggregation for accuracy and execution time are outlines, and the method for measuring uncertainty is introduced.

\section{Disaggregation with Neural Networks}
\label{sec:disag_w_nn}
The primary goal of this project was to implement disaggregation using neural networks. This was achieved using the Keras \citep{keras} package in R as an interface into TensorFlow.

\subsection{Neural Network Disaggregation Model Architecture}
This section will describe the architecture of a neural network model that has been designed to implement disaggregation regression.

Consider only one region: we have a vector with the geometry describing the boundary and a value for the number of cases reported within that region. There is also a stack of rasters that has been cropped to the region’s boundary, which contains the data for each covariate at every pixel within the region. An additional raster contains the population count in each pixel at the same resolution as the covariates.

In the simplest case (a GLM), the covariates of each pixel can predict the rate \(\mu\) at that pixel and using the population \(p\) as an offset, the number of cases \(y\) can be calculated:

\begin{equation}
    \eta_i=\beta_0+x_{1i}\beta_1+\cdots
\end{equation}

\begin{equation}
    \mu_i=g^{-1}\left(\beta_0+x_{1i}\beta_1+\cdots\right)
\end{equation}

\begin{equation}
    y_i=\mu_ip_i
\end{equation}

Expressed as matrices, where each row of \(\eta\) represents the linear predictor at each pixel within the region:

\begin{equation}
    \left[\begin{matrix}\eta_0\\\eta_1\\\vdots\\\end{matrix}\right]=\left[\begin{matrix}1&x_{11}&\ldots\\1&x_{21}&\ldots\\\vdots&\vdots&\ddots\\\end{matrix}\right]\left[\begin{matrix}\beta_0\\\beta_1\\\vdots\\\end{matrix}\right]
\end{equation}

\begin{equation}
    \left[\begin{matrix}\mu_1\\\mu_2\\\vdots\\\end{matrix}\right]=g^{-1}\left(\left[\begin{matrix}\eta_0\\\eta_1\\\vdots\\\end{matrix}\right]\right)
\end{equation}

\begin{equation}
    \left[\begin{matrix}y_1\\y_2\\\vdots\\\end{matrix}\right]=\left[\begin{matrix}\mu_1\\\mu_2\\\vdots\\\end{matrix}\right]\circ\left[\begin{matrix}p_1\\p_2\\\vdots\\\end{matrix}\right]
\end{equation}

Alternatively:

\begin{equation}
    \boldsymbol{H}=\boldsymbol{X}\boldsymbol{B}
\end{equation}

\begin{equation}
    \boldsymbol{M}=g^{-1}\left(\boldsymbol{H}\right)
\end{equation}

\begin{equation}
    \boldsymbol{Y}=\boldsymbol{M}\circ\boldsymbol{P}
\end{equation}

In disaggregation, only the total cases in the region are known, which is the sum of the column matrix \(\boldsymbol{Y}\). This can be calculated by taking the dot product of the rate \(\boldsymbol{M}\) and population \(\boldsymbol{P}\), instead of the Hadamard product:

\begin{equation}
    y=\boldsymbol{M}\cdot\boldsymbol{P}
\end{equation}

\begin{figure}
    \centering
    \includegraphics{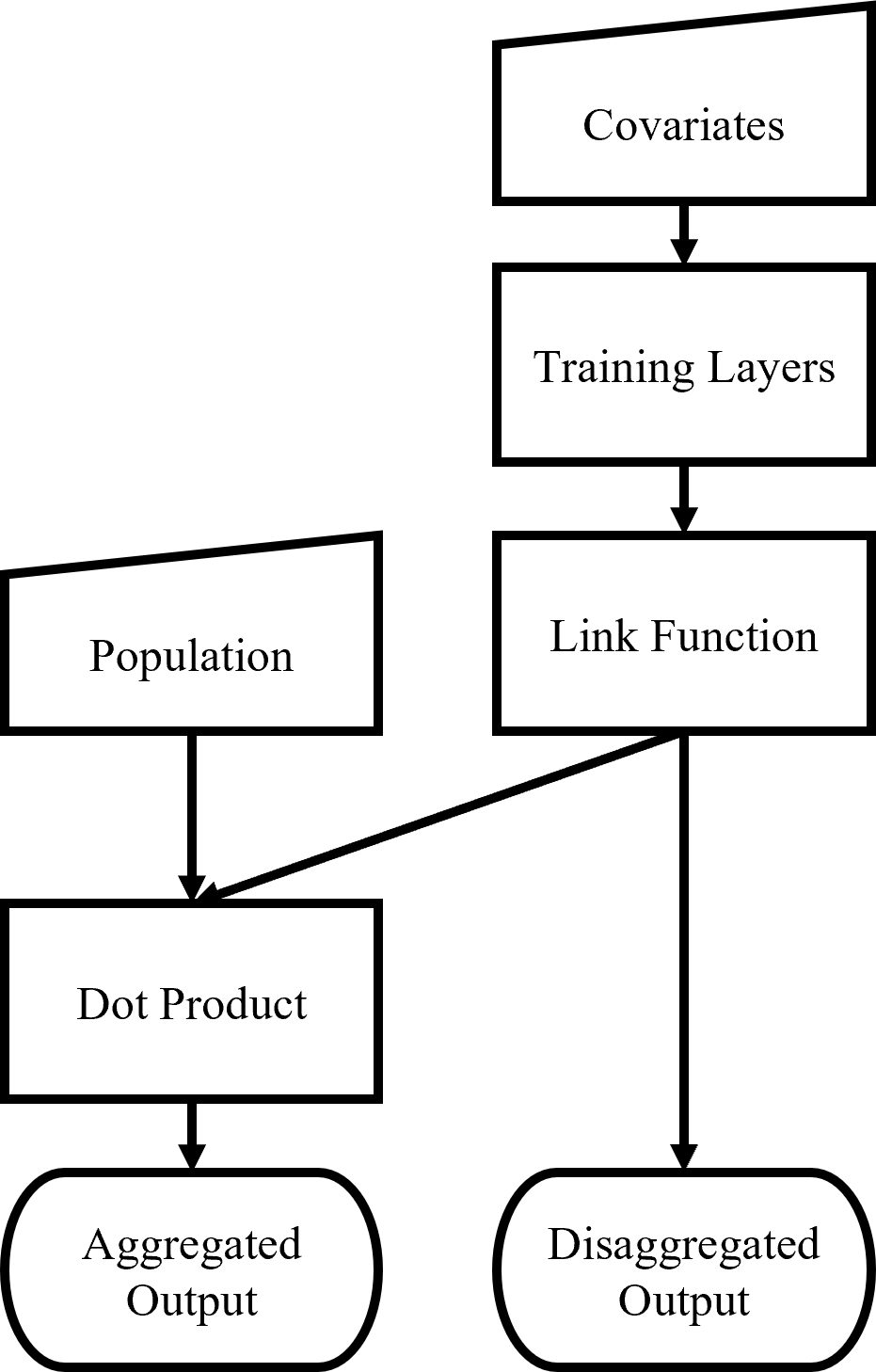}
    \caption{Directed Acyclic Graph of a Disaggregation Neural Network model}
    \label{fig:dag1}
\end{figure}

The linear predictor part of the equation, \(\boldsymbol(H)\), can have more complexity than a linear regression. A neural network can be used to take the input variables \(\boldsymbol{X}\) and return a linear predictor \(\boldsymbol{H}\) using a learned relationship of any complexity. We are interested in finding the column matrix \(\boldsymbol{M}\), which represents the rate of disease for each pixel. This can be extracted from the relevant layer in a neural network after training (Figure~\ref{fig:dag1}).

In Figure~\ref{fig:dag1}, the input “Covariates” would be a 2d tensor, with one dimension being each pixel and the other being the number of layers (each layer as a covariate). Population would also be 2d, although only with one layer. Training layers train per pixel; for an input with 1000 pixels and 4 covariates, the training layers would only be concerned with the 4 layers per pixel, so the weights for each pixel would be the same. Keras provides this functionality for tensors of any dimension, as densely connected layers will only consider the ‘first’ on any number of dimensions. The final training layer must be a densely connected layer with one unit and a linear activation function. This would represent the linear predictor portion of a Generalised Linear Model and has a value for each input pixel. Applying the link function gives the disaggregated output that we are interested in. The ‘aggregated output’ is the value which the model is trained to. Each region is therefore considered one data point, and the model is trained on multiple regions.

\begin{figure}[!h]
    \centering
    \includegraphics[width=\textwidth]{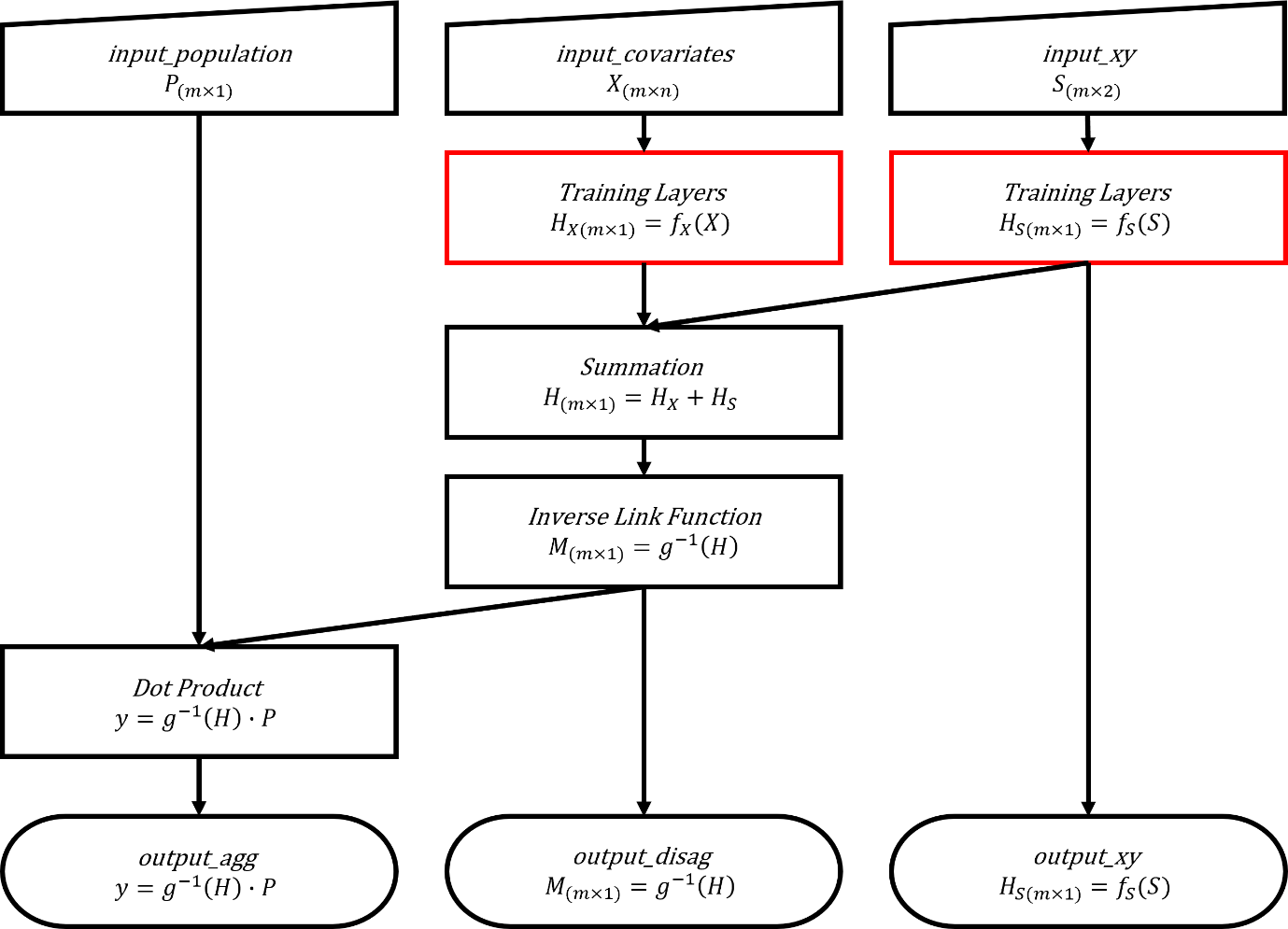}
    \caption{Directed Acyclic Graph of the full neural network disaggregation model}
    \label{fig:dag2}
\end{figure}

Figure~\ref{fig:dag2} outlines the full model that was designed for disaggregation with neural networks. The mathematical symbols and functions used in the model are:
\begin{itemize}
    \item \(\boldsymbol{P}_{(m\times1)}\): \verb|input_population|. An \(m\times1\) tensor where m is the number of pixels and each element is the population in that pixel.
    \item \(\boldsymbol{X}_{(m\times n)}\): \verb|input_covariates|. An \(m\times n\) tensor where m is the number of pixels and n is the number of covariates. Each element is the normalised value for the relevant covariate and pixel.
    \item \(\boldsymbol{S}_{(m\times2)}\): \verb|input_xy|. An \(m\times2\) tensor where m is the number of pixels and has two columns, one for latitude and longitude. The values are normalised.
    \item \(\boldsymbol{H}_{X(m\times1)}=f_X(\boldsymbol{X})\): Training layers for the covariates. \(f_X\) applies row-wise and returns an \(m\times1\) tensor, where each row is the covariate portion of the linear predictor.
    \item \(\boldsymbol{H}_{S(m\times1)}=f_S(\boldsymbol{S})\): Training layers for the spatial data. \(f_S\) applies row-wise and returns an \(m\times1\) tensor, where each row is the spatial portion of the linear predictor. This is returned to the user as \verb|output_xy|, which can be used to plot the effect of the spatial field.
    \item \(\boldsymbol{H}_{(m\times1)}=\boldsymbol{H}_X+\boldsymbol{H}_S\): The sum of the covariate and spatial parts of the linear predictor. Returns an \(m\times1\) tensor which is the full linear predictor for each pixel.
    \item \(\boldsymbol{M}_{(m\times1)}=g^{-1}(\boldsymbol{H})\): The inverse link function, applied element wise, e.g. with a Poisson distribution, the inverse link function would be an exponential, and the output would be an \(m\times1\) tensor of disease rates in each pixel. This is returned to the user as \verb|output_disag| and is the primary output of interest.
    \item \(y=g^{-1}\left(\boldsymbol{H}\right)\cdot \boldsymbol{P}\): The dot product of the tensor of rates and population. Multiplies each pixel by the population to get a disease count, then sums across the entire region. Returns a scalar, y, which is the total disease count across the whole region, as \verb|output_agg|.
\end{itemize}

\subsection{Data Pre-Processing}

Keras, when used with R, takes input data as arrays (also referred to as tensors). Spatial data is represented as vectors or rasters, so pre-processing and formatting is necessary to convert the data to a format recognised by Keras. Multiple inputs in the Keras R package must be represented as a list of arrays, and the dimensions of each data point must be the same. Regions rarely have the same number of pixels, so the size of the input arrays must be the size of the largest region; smaller regions can be padded with zeros to the correct size as the dot product will effectively ignore pixels with a population of zero. The Madagascar dataset has 4 covariates, 109 regions, and the largest region is 867 pixels: the dimensions of the covariate input array is therefore \verb|(109, 867, 4)| and the dimensions of the population input array is \verb|(109, 867, 1)|. The aggregated output is a vector of length 109, with each element representing the total cases in each region.

\subsection{Training}
As the disaggregated output is not known, training takes place using the aggregated output as the truth, and the disaggregated output is ignored during the training process. The optimiser aims to reduce the Poisson loss between the count of cases predicted by the model and the count of cases supplied as truth, by adjusting the weights in the training layers that represent the relationship between the covariates. As there are only 109 regions, training must take place over a high number of epochs (repeats); the actual number of epochs depends on the optimiser, the designer, and any early stopping criteria.

Validation splits may be used to reduce variance. In this case, a subset of the original dataset is held out and not used for training. After each epoch, the accuracy of the model is measured against the held out (validation) dataset and is reported alongside the in-sample accuracy to offer the designer insight into whether the model has bias or variance. Ideally, the validation accuracy would be close to the training accuracy, as this represents a model that fits well to the training data as well as unseen data.

\subsection{Prediction}

Prediction with a neural network is made by simply introducing a new dataset to the model and recording the results. Disaggregation neural networks use the same method, although some adjustments must be made to the model to account for the non-typical architecture.

Recall that the full covariate raster set has been divided into subsets by the borders of the regions in the vector polygons, then 'straightened' and padded with zeros to create a set of arrays of the same size. For predictions to take place, new covariate rasters must be reshaped into this format and match the dimensions. Taking the Madagascar dataset with covariate array dimensions of \verb|(109, 867, 4)| and the dimensions of the population input array being \verb|(109, 867, 1)|, dimensions for the prediction data should match this. This is straightforward, as no specific borders are necessary. The stack of rasters is converted into a dataframe where each row represents a pixel and each column represents a layer, then the dataframe is split into sets of \verb|867|. Only the last set will require zero padding.

The final predicted values from the disaggregated output can be concatenated with the coordinate data to rebuild a final raster of predictions.

If the covariates used for training have been normalised and scaled, then this should be considered when making predictions. The prediction data should be scaled and normalised to the mean and variance of the training data, otherwise predictions would be incorrect.

\section{The 'Kedis' R Package}
\label{sec:kedis}
The implementation of disaggregation with neural networks has been converted into a package in R called \textit{Kedis} (Keras-Disaggregation) \citep{kedis}, which can be accessed at \url{github.com/jackahall/kedis}. It provides a user-friendly interface to implement disaggregation using neural networks in production. For a user familiar with the Keras API it is straightforward to use. The package can be installed using the following commands (importing the Kedis library will also import the Keras library, and requires an installation of Python and TensorFlow):

\begin{verbatim}
remotes::install_github("jackahall/kedis")
library(kedis)
\end{verbatim}

Data can then be read into the environment (\verb|covariates.tif| is a raster file that includes all covariates of the model; they can be imported individually and stacked manually).

\begin{verbatim}
library(terra)
shapes <- vect("data/shapes/mdg_shapes.shp")
covatiates <- rast("data/covariates.tif")
population <- rast("data/population.tif")
\end{verbatim}

The function \verb|prepare_data| takes the shapefile, covariates and population rasters and builds a \verb|kd_data| object.

\begin{verbatim}
data <- prepare_data(shapes = shapes,
                     covariates = covariates,
                     population = population,
                     filter_var = "ID_2",
                     response_var = "inc")
\end{verbatim} 

Next, lists of hidden layers for covariates and coordinates are specified. These can be any Keras layer type, such as dense, dropout or Gaussian noise, and this is a major strength in terms of flexibility of this method.

\begin{verbatim}
layers_cov <- list(layer_dense(units = 20, activation = "relu"),
                   layer_dropout(rate = 0.2),
                   layer_dense(units = 10, activation = "relu"))
                   
layers_xy <- list(layer_dense(units = 20, activation = "relu"),
                  layer_dropout(rate = 0.2),
                  layer_dense(units = 10, activation = "relu"))
\end{verbatim}

Using these layers, a Kedis model is built. For those familiar with Keras, options for the compiler can also be supplied here. If a loss function is not supplied, then the model will not be compiled and must be compiled manually.

\begin{verbatim}
model <- build_model(data = data,
                     layers_cov = layers_cov,
                     layers_xy = layers_xy,
                     link = "log",
                     optimizer = optimizer_adam(),
                     loss = loss_poisson())
\end{verbatim}

Finally, the model can be trained. Additional arguments will be passed through to \verb|keras::fit|, such as \verb|epochs|, \verb|validation_split| and \verb|callbacks|.

\begin{verbatim}
history <- train(model,
                 epochs = 1000,
                 validation_split = 0.2)
\end{verbatim}

The \verb|history| object inherits from the Keras history class, and also includes execution time. Plotting \verb|history| works in the same way as plotting Keras training histories.

The disaggregated prediction raster can be generated using \verb|predict|, which returns a \verb|SpatRaster| of the prediction:

\begin{verbatim}
prediction <- predict(model)
\end{verbatim}

Or alternatively, plotted directly:

\begin{verbatim}
plot(model)
\end{verbatim}

The package also includes functions to perform cross-validation, obtain various loss and model metrics (such as mean absolute error or root mean squared error), and to make predictions using a new dataset.

\section{Direct Comparison of Simple Models with no Spatial Effect}
\label{sec:direct_comp}

A Kedis model with no additional hidden layers should predict the same output as a disaggregation model with no field or IID effect. This straightforward 1-1 test was used to determine whether the Kedis algorithm was working as expected.

A disaggregation model with the field and IID effect turned off was fitted. A Kedis model with no hidden layer was built and trained to a point where the loss change no longer had a downward trend (the model will jitter around a minimum, but the trend will stop). At this point, the performance of Kedis was compared to disaggregation by:

\begin{enumerate}
    \item Comparing the weights. A Kedis model with no hidden layers and four covariates will have five weights (one per covariate, plus a bias), which should match closely to the coefficients of the disaggregation model.
    \item Visually comparing the prediction plots. The models should produce similar maps, with areas of high and low incidence matching.
    \item Reaggregating the predictions and comparing the incidence rate of each region with the true value, to look for trends in the predictions or bias.
\end{enumerate}

It is not expected that the models will match exactly, for a few reasons:

\begin{enumerate}
    \item Disaggregation works on a Bayesian framework and includes priors in its estimation, whereas Kedis works only with the dataset provided
    \item The optimisation of Kedis will not completely settle on a minimum, but will move around that point.
    \item The Kedis optimiser may settle on a local minimum.
\end{enumerate}

The purpose of this test is not to exactly replicate disaggregation with Kedis, but rather to ensure the model is behaving in a stable and expected manner. Small differences in predictions are expected and acceptable.

\section{Cross-Validation of Disaggregation and Kedis}
Comparison between ‘traditional’ disaggregation and the neural network implementation was done using a repeated cross-validation on each method. A k-fold cross-validation splits the dataset into \(k\) subsets, and a model is fitted leaving out each fold in turn to be used as a validation set. For each model, the validation set is used to assess the out-of-sample accuracy of the fitted model. The ‘final’ model accuracy is the average of the \(k\) models.

Cross-validation is a method to measure the predictive accuracy of a model, as the validation data is not seen during training. This reduces the variance (overfitting), as an overfitted model would not perform well when introduced to previously unseen data.

Repeated cross-validation takes the idea of cross-validation further, by repeating this method with the data split differently each time. The scores from each model are averaged, and this gives a more robust assessment than running a cross-validation only once. As each repeat uses a different data split, the average accuracy scores will be different, leading to a distribution of scores around the true average.

In cross-validating disaggregation regression methods, the data is split by regions, not by individual pixels. The Madagascar dataset comprises of 109 regions, so in a 5-fold cross validation this would be split into 5 folds of approximately 22 regions each. Stratified cross validation takes the data split beyond random assignment and aims to ensure that each fold is representative of the overall dataset. In a regression cross validation (as opposed to classification), this is achieved by ensuring the mean and standard deviation of each fold is approximately that of the full dataset. As each test fold has a response similar to the test data, this method avoids issues around extrapolation. In R, stratified data splitting can be easily achieved using the \verb|createFolds| function in the ‘caret’ package \citep{caret}. Stratified cross-validation was performed as standard and will hereinafter be referred to simply as cross-validation.

\begin{figure}
    \centering
    \includegraphics{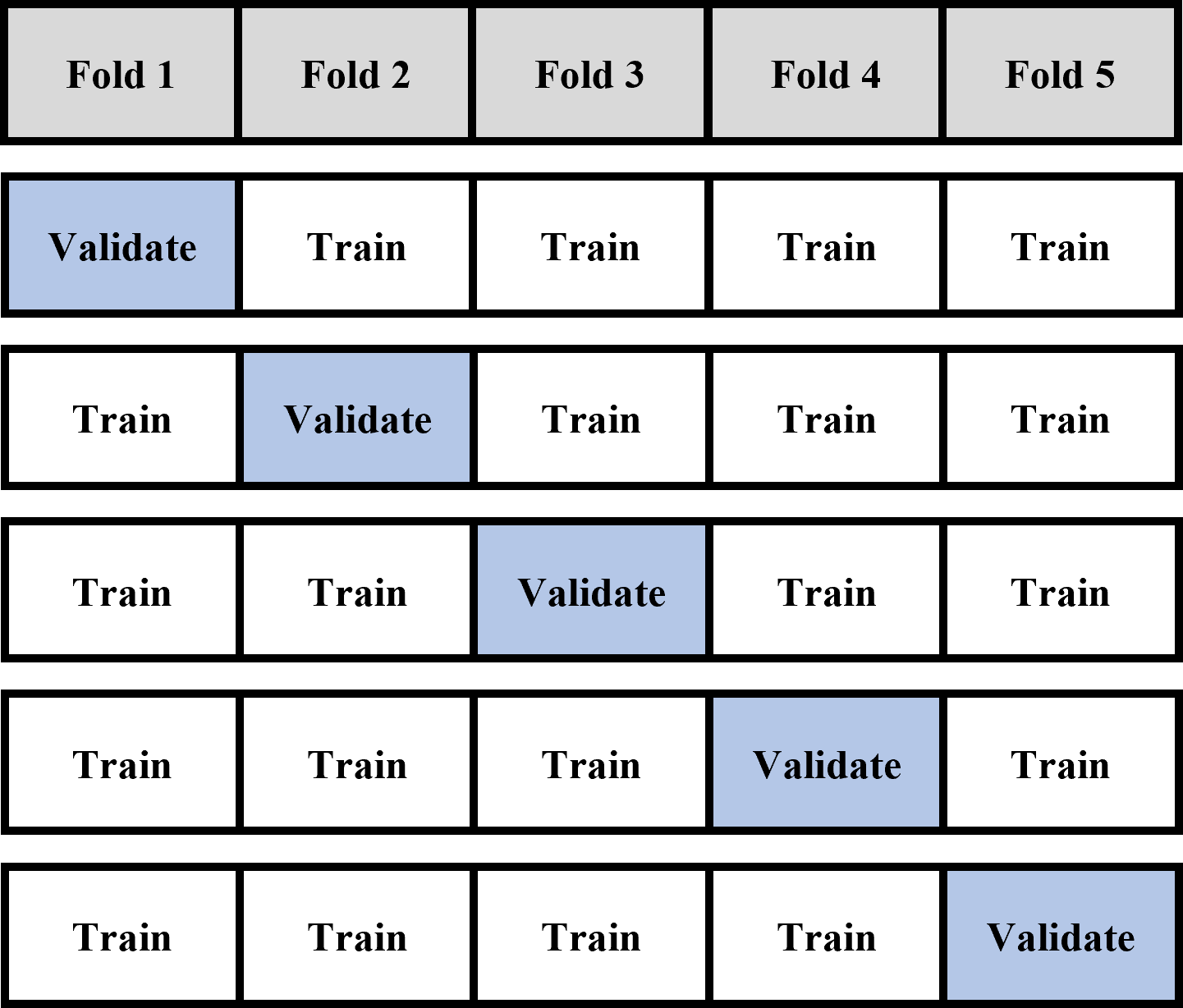}
    \caption{Diagram of Cross-Validation split}
    \label{fig:cv}
\end{figure}

Using the Madagascar malaria dataset, cross validation was performed on disaggregation and Kedis. In total, six models were examined:

\begin{enumerate}
    \item Disaggregation with field and IID turned off
    \item Disaggregation with field and IID turned on
    \item Kedis with no spatial data
    \item Kedis with latitude and longitude as covariates
    \item Kedis with one dense layer (100 nodes) for spatial analysis
    \item Kedis with one dense layer (1000 nodes) for spatial analysis
\end{enumerate}

Each cross-validation was repeated twenty times, with a different data split for each repeat. The same random seed was used for each model to ensure the set of data splits was the same for each model.

The Kedis models were trained on 10,000 epochs, with an early stopping callback monitoring the validation loss, with a minimum delta of \(1\times10^{-10}\) and a patience of 10. This was achieved using a \verb|callbacks| option in Keras:

\begin{verbatim}
callback_early_stopping(monitor = "val_loss",
                        min_delta = 1e-10,
                        patience = 10)
\end{verbatim}

The distribution of losses was compared using boxplots, then t-tests and ANOVA were used to determine statistical significance if necessary.

\section{Hyperparameter Selection with Nested Cross-Validation}
Hyperparameters are variables of the model that are not learned during training. In artificial neural networks, examples of these could be number of nodes in each layer, number of layers, or activation function. Model performance is dependent on the particular hyperparameters selected: a model with more layers and nodes tends to overfit (higher variance) more than one with few layers and nodes, which would have a high bias; activation function choice may have an unpredictable effect on the model, as can any number of possible hyperparameters. For any dataset, there are one or more sets of hyperparameters that best balance the bias-variance trade-off and give optimal performance. The challenge for a developer is finding a set of hyperparameters that work well with a given dataset; nested cross-validation is one method to achieve this.

The algorithm for nested cross-validation can be explained in pseudocode in Algorithm~\ref{alg:ncv}.

\begin{algorithm}
\caption{Nested Cross-Validation} \label{alg:ncv}
\begin{algorithmic}

\STATE Define a set of hyperparameters, $C$
\STATE Divide dataset into $K_{OUTER}$ folds

\FOR{fold $k_{out}$ in $K_{OUTER}$}
    \STATE Set fold $k_{out}$ as test set
    \STATE Set remaining $K_{OUTER}-1$ folds as outer train set
    \STATE Divide outer train set into $K_{INNER}$ folds
    
    \FOR{hyperparameter set $c$ in $C$}
    
        \FOR{fold $k_{in}$ in $K_{INNER}$}
            \STATE Set fold $k_{in}$ as validation set
            \STATE Set remaining $K_{INNER}-1$ folds as inner train set
            \STATE Train model on inner train set
            \STATE Evaluate performance on validation set
        \ENDFOR
        \STATE Calculate average performance of hyperparameter set $c$
    \ENDFOR
    \STATE Take best performing hyperparameter set $c$
    \STATE Evaluate performance on test set
\ENDFOR
\STATE Calculate average performance over each $K_{OUTER}$ folds

\end{algorithmic}
\end{algorithm}

For each outer fold, it is possible that the best hyperparameter set will be different. Nested cross-validation does not assess the performance of an ‘exact’ model, rather the overall performance of the architecture and data. 

A 5-outer-5-inner-fold nested cross-validation was conducted on Kedis, with 207 sets of hyperparameters. The hyperparameter sets only included layers for covariate training, and hyperparameters for the optimiser were all default. Number of nodes in a dense layer, the number of dense layers, and dropout with different rates was assessed.

The hyperparameter set was comprised of stacks of a dense layer followed by a dropout layer. Up to four stacks were assessed. Each dense layer had between 2-20 nodes and a ReLU activation, and each dropout layer had a rate of either 0, 0.1 or 0.2 (where there was more than one dropout layer, the rate was the same in each). As fully assessing the full set of increasingly deep models is computationally inefficient (\(\mathcal{O}(n^2)\)), a random search was conducted from the full set of hyperparameters. Table~\ref{tab:hypers} demonstrates how the full set of 207 hyperparameters was chosen. The nested cross-validation was performed twice: with and without coordinates as covariates. The same seed was set for each of the two models, so the hyperparameter sets and starting values in Keras were the same.

\begin{table}[H]
\centering
\begin{tabular}{lcccc}
  \toprule
Depth & Nodes & Rate & Combinations & Selected \\ 
  \midrule
1 & 2-20 & 0, 0.1, 0.2 & 57 & 57 \\
2 & 2-20 & 0, 0.1, 0.2 & 1,083 & 50 \\
3 & 2-20 & 0, 0.1, 0.2 & 20,577 & 50 \\
4 & 2-20 & 0, 0.1, 0.2 & 390,963 & 50 \\
\addlinespace
\multicolumn{3}{l}{Total Combinations} & 412,680 & 207 \\
\bottomrule
\end{tabular}
\caption{\label{tab:hypers} Out-of-sample losses for 5-fold repeated cross-validation on six models}
\end{table}

\section{Execution Time Comparison}

The execution time of fitting six models were compared (the same models investigated in the Cross-Validation section):

\begin{enumerate}
    \item Disaggregation with field and IID turned off
    \item Disaggregation with field and IID turned on
    \item Kedis with no spatial data
    \item Kedis with latitude and longitude as covariates
    \item Kedis with one dense layer (100 nodes) for spatial analysis
    \item Kedis with one dense layer (1000 nodes) for spatial analysis
\end{enumerate}

The data preparation steps and the model fitting steps were timed separately. The Kedis models had \verb|kedis::prepare_data()| measured as the data processing step, and \verb|kedis::build_model()| and \verb|kedis::train()| functions measured together as the model fitting step. Each model was built and fitted 50 times. The mean execution time of fitting each model was compared.

The Kedis model training time is heavily dependent on the optimiser and number of epochs. For the training time comparison, each model used an Adam optimiser, which according to the authors \citep{kingma2014} is computationally efficient, has little memory requirement, is invariant to diagonal re-scaling of gradients, and is well suited for problems that are large in terms of data/parameters. The default parameters set by Keras \citep{chollet2015keras} were used, although these could possibly be optimised further for faster training. The number of epochs was set at 10,000, with an early stopping callback with parameters:

\begin{itemize}
    \item \verb|monitor = "loss"|
    \item \verb|min_delta = 1e-10|
    \item \verb|patience = 50|
\end{itemize}

The distribution of execution times was compared using boxplots, then t-tests and ANOVA were used to determine statistical significance if necessary.

\section{Uncertainty Estimates with Monte Carlo Dropout}

Monte Carlo dropout was first introduced by \citet{gal2016} as a method of exploiting dropout layers to estimate model uncertainty. The traditional use of a dropout layer is to improve the generalisability of a model by setting to zero the output nodes at random during training time. This behaviour makes it more difficult for the model to simply remember the dataset as the weights will be spread more evenly across all the hidden nodes. Dropout layers are only used during training time; during inference these layers are ignored, so predictions are always the same. Monte Carlo dropout changes the action off hidden dropout layers to be active during inference, thus returning a different prediction each time, then sampling repeatedly. The range of outputs will be used to estimate the overall uncertainty of a model, and can also improve the accuracy of mean predictions \citep{oleszak_2021}. 

The feasibility of Monte Carlo dropout in disaggregation regression with neural networks was assessed with a simple design study. A minor change was required to the \verb|build_model| function in the Kedis package, forcing layers to believe they are undergoing training which has the effect of activating dropout during inference. A simple Kedis model was built with a single hidden dropout layer and trained over 10,000 epochs with an early stopping criteria. Estimates of uncertainty were inferred from the distribution of the predictions at each pixel.

\chapter{Results}
\label{ch:results}

\section{Direct Comparison of Simple Models with no Spatial Effect}
\label{sec:direct_comp_res}

The simplest possible Kedis model is one with no hidden layers and fits a linear combination of covariates with no spatial effect. This is the same model that is fitted by traditional disaggregation regression with the field and IID turned off, so each models should return similar results. Table~\ref{tab:direct} outlines the weights for each covariate in the two models. With the exception of elevation all parameters are the same sign, although there are large differences in magnitudes of every parameter except intercept.

\begin{table}[H]
\centering
\begin{tabular}{lcc}
  \toprule
Variable & Disaggregation & Kedis \\ 
  \midrule
(intercept) & -2.92808 & -2.59353 \\ 
\addlinespace

  Elevation & -0.69121 & 0.20678 \\ 
  EVI & 0.03858 & 0.34690 \\ 
  LST\_mean & 0.16541 & 0.78791 \\ 
  LST\_sd & -0.09824 & -0.20189 \\  
   \bottomrule
\end{tabular}
\caption{\label{tab:direct} Parameters calculated by the simple disaggregation and Kedis models}
\end{table}

\begin{figure}[H]
    \centering
    \includegraphics[width=\textwidth]{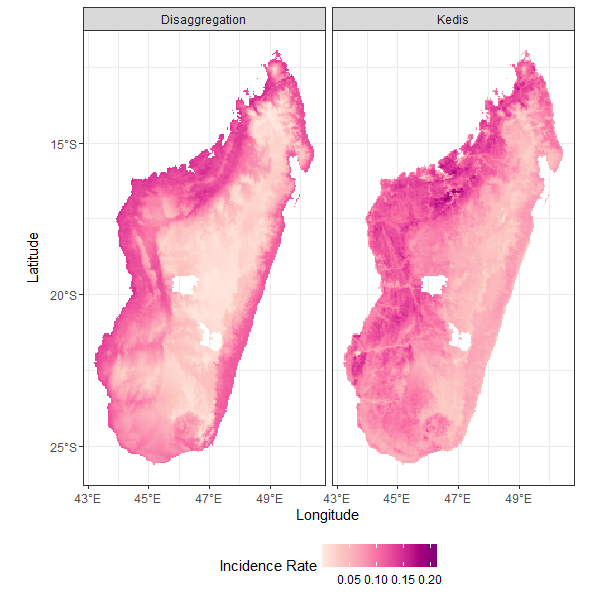}
    \caption{Maps showing the disaggregated predictions generated by disaggregation with no field or IID, vs a Kedis model with no hidden layers.}
    \label{fig:direct}
\end{figure}

Figure~\ref{fig:direct} displays the  maps of incidence predictions created by each model. A visual inspection shows the two models predict a similar pattern in malaria incidence, albeit with some differences in magnitudes. Both models show a higher incidence around the west coast of Madagascar and an additional narrow region of higher incidence along the east coast, with the central regions being of lowest incidence. Close inspection shows the same small local patterns in most areas, such as in the very north and in the south-east, which suggests the Kedis model is performing a similar calculation to disaggregation.

\begin{figure}[H]
    \centering
    \includegraphics[width=\textwidth]{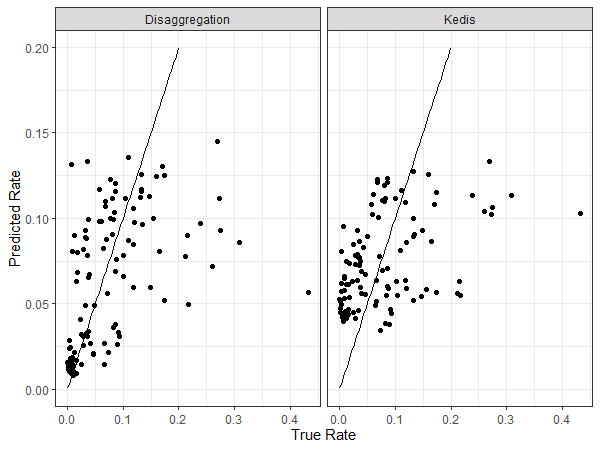}
    \caption{Scatter plot of true rates vs rates predicted in-sample by disaggregation with no field or IID and a Kedis model with no hidden layers. Black line is an identity line and represents perfect predictions.}
    \label{fig:direct_scatter}
\end{figure}

Figure~\ref{fig:direct_scatter} shows the rates for each of the 109 regions in Madagascar that were predicted by each model, against the true rates. These values are calculated as the sum of the counts in each pixel for a given region, divided by the sum of the population in each pixel for the same region. These plots show that the Kedis model does not predict any regions to have a rate below around 0.03, and the range of predictions is a lot lower than than actual data. Disaggregation predictions match the true rates where the incidence is low better than Kedis, although both models perform badly in predicting the regions with the highest true rates (demonstrated by the high spread to the right of the identity line). 

\begin{table}[H]
\centering
\begin{tabular}{lccccc}
  \toprule
Model & n & Mean & SD & Min & Max \\ 
  \midrule
Actual & 109 & 0.07936 & 0.07914 & 0.00049 & 0.43248 \\ 
\addlinespace
  Disaggregation & 109 & 0.06525 & 0.04018 & 0.00840 & 0.14476 \\ 
  Kedis & 109 & 0.07449 & 0.02633 & 0.03426 & 0.13348 \\ 
   \bottomrule
\end{tabular}
\caption{\label{tab:direct_summary} Summary statistics of predicted and actual rates for disaggregation and Kedis models over every region in Madagascar.}
\end{table}

Table~\ref{tab:direct_summary} outlines the summary statistics of the data displayed in Figure~\ref{fig:direct_scatter}. It shows that although both models are good at predicting the mean, the actual standard deviation and range is higher in the actual data than predicted by both models. 

\begin{figure}[H]
    \centering
    \includegraphics[width=\textwidth]{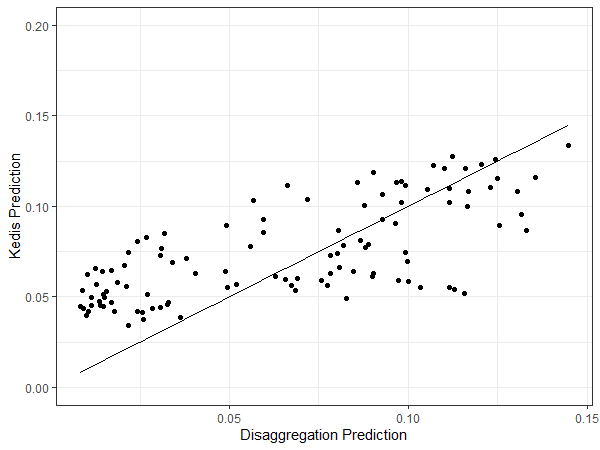}
    \caption{Scatter plot of predictions made by the disaggregation model with no field or IID, vs predictions made by Kedis with no hidden layers. Correlation = 0.71. Black line is an identity line and represents perfect agreement.}
    \label{fig:cor_disag_kedis}
\end{figure}

The correlation between predictions made by disaggregation and Kedis was 0.71, showing a fair agreement between the two models, and is displayed in Figure~\ref{fig:cor_disag_kedis}. The overestimation of lower incidence points by Kedis can be seen here, as every point on the left hand side of the plot is estimates to be higher by Kedis than disaggregation.

\begin{figure}[H]
    \centering
    \includegraphics[width=\textwidth]{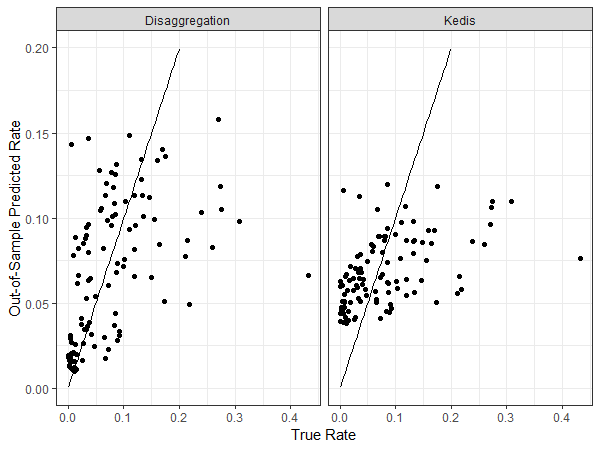}
    \caption{Scatter plot of true rates vs out-of-sample predictions made by a disaggregation model with no field or IID and a Kedis model with no hidden layers. Black line is an identity line and represents perfect predictions.}
    \label{fig:cv_scatter}
\end{figure}

Figure~\ref{fig:cv_scatter} shows the out-of-sample predictions for a simple disaggregation model and a simple Kedis model against the true rate. The pattern observed here is similar to the pattern of in-sample predictions made in Figure~\ref{fig:direct_scatter}, where Kedis overestimates small values and both models do a poor job at estimating larger values. The Pearson's product moment correlation coefficient \(r\) for both models is almost identical (disaggregation = 0.494, Kedis = 0.490), although a high correlation may not correspond to a good prediction model if the relationship does not follow the identity line.

\section{Cross Validation Results}
\label{sec:cv_res}

The 5-fold cross-validation was performed on six unique models:

\begin{enumerate}
    \item Disaggregation - Field/IID off
    \item Disaggregation - Field/IID on
    \item Simple Kedis: a basic Kedis model with no hidden layers
    \item Kedis - xy as covariates: a basic Kedis model with no hidden layers, but with coordinates included as covariates
    \item Kedis - 100 units for space: a Kedis model with a single dense layer with 100 units and ReLU activation on the coordinate training side
    \item Kedis - 1000 units for space: a Kedis model with a single dense layer with 1000 units and ReLU activation on the coordinate training side
\end{enumerate}

\begin{figure}
    \centering
    \includegraphics[width=\textwidth]{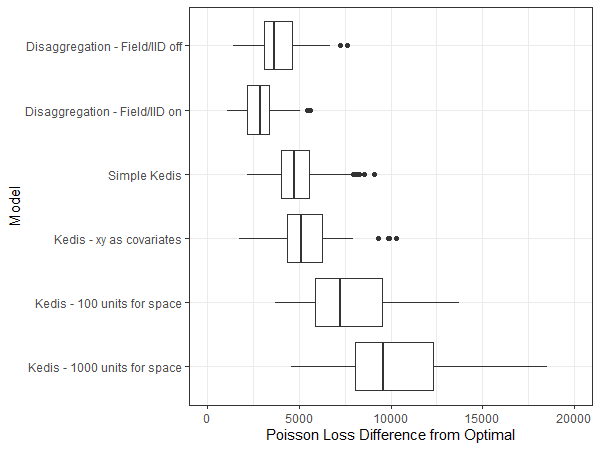}
    \caption{Boxplot of the repeated cross validation results}
    \label{fig:cv_box}
\end{figure}

Each 5-fold cross-validation was repeated 20 times, each with a different data split, which gives 100 values for out-of-sample accuracy. The summary for these losses is outlined in Table~\ref{tab:cv} and displayed in a boxplot in Figure~\ref{fig:cv_box}.

\begin{table}[H]
\centering
\begin{tabular}{lccccc}
  \toprule
Model & N & Mean & SD & \multicolumn{2}{c}{95\% CI} \\ 
  \midrule
Disag - Field/IID off & 100 & 3843 & 1159 & 3613 & 4073 \\ 
  Disag - Field/IID on & 100 & 2852 & 943 & 2665 & 3039 \\ 
  \addlinespace
  Kedis - xy off & 100 & 5042 & 1556 & 4733 & 5350 \\ 
  Kedis - xy on & 100 & 5355 & 1648 & 5028 & 5682 \\ 
  Kedis - xy 100 nodes & 100 & 7824 & 2790 & 7270 & 8377 \\ 
  Kedis - xy 1000 nodes & 100 & 10216 & 3307 & 9560 & 10872 \\ 
\bottomrule
\end{tabular}
\caption{\label{tab:cv} Out-of-sample losses for 5-fold repeated cross-validation on six models}
\end{table}

An ANOVA on the six groups confirms there is a statistically significant difference between at least two of the groups (F = 169.556, p $<$ 0.00001). After excluding the two groups with a dense layer on the coordinate training side (they are visually different to the other groups), the difference between the groups is still very significant (F = 71.742, p $<$ 0.00001). Any adjustment for multiple testing would make a negligible impact on levels of significance in this case.

The worst performing disaggregation model (field and IID turned off) had significantly lower loss than the best performing Kedis model (simple with no hidden layers of coordinates as covariates) when assessed with a t-test ($\mu$: -1198.234; 95\% CI: [-1581.0177, -815.4517]; p $<$ 0.00001), and a conservative Bonferroni correction (assuming \(6^2=36\) comparisons reducing \(\alpha\) to 0.001389) made no change to significance ($\mu$: -1198.234; AdjCI: [-1828.1158, -568.3535], p $<$ 0.00001). It can confidently be stated that the out-of-sample accuracy of Kedis is worse than that of disaggregation.

The assumptions of the ANOVA and t-test were assessed, and the results confirmed with non-parametric Kruskal-Wallis test and Mann-Whitney test respectively.

\section{Kedis Hyperparameter Tuning}

Two nested cross-validation were performed with Elevation, Enhanced Vegetation Index (EVI), Land Surface Temperature Mean and Land Surface Temperature Standard Deviation included as covariates in both cases, and Latitude and Longitude (normalised) included in one. The 109 regions were split into five outer folds, and each inner training fold was split again into five folds. 207 combinations of covariate training layers were fitted. 

\begin{figure}
    \centering
    \includegraphics[width=\textwidth]{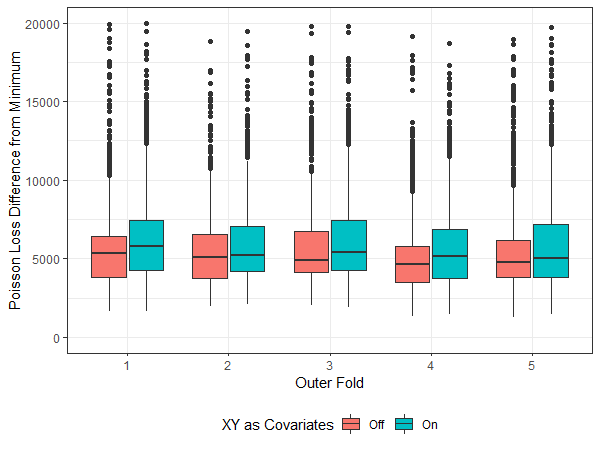}
    \caption{Boxplot showing the losses for each inner loop of the nested cross-validation, with space turned off and on. N.B. 38 data points have been cropped as they had values over 20,000, but were evenly distributed across each group.}
    \label{fig:ncv_inner}
\end{figure}

\begin{table}[H]
\centering
\begin{tabular}{rrr}
  \toprule
Outer Loop & XY Off & XY On \\ 
  \midrule
  1 & 5744.87 & 6376.79 \\ 
    2 & 5608.21 & 6014.45 \\ 
    3 & 5863.28 & 6385.52 \\ 
    4 & 5201.15 & 5742.92 \\ 
    5 & 5623.01 & 5960.86 \\ 
    \midrule
    Mean & 5608.10 & 6096.11 \\ 
   \toprule
\end{tabular}
\caption{\label{tab:ncv_inner} Mean losses for the nested cross-validation for each inner group, with the overall mean for the inner folds}
\end{table}

Figure~\ref{fig:ncv_inner} outlines the losses on the inner folds for the nested cross-validation runs. Table~\ref{tab:ncv_inner} shows the means of the same data. Based on the inner folds, including space as a covariate has a detrimental effect on the accuracy of the model (Loss = 5608.10 for no spatial covariate, vs 6096.11 for spatial covariate). Given that these models were fitted using the same hyperparameter set, and the pattern is similar across each of the five folds, this is a fair comparison despite not being the primary outcome measure of nested cross-validation.

The best performing hyperparameter sets for each inner fold were refitted using the relevant four outer folds, and tested against the fifth remaining fold. The results of this final step is outlined in Table~\ref{tab:ncv_outer}. The mean of the 'outer loss' column is the mean of the outer folds, and is the primary measure of accuracy of a model assessed using nested cross-validation. With space turned off, the model has a loss of 7210.53; with space turned on, the model has a loss of 4252.36. 

Due to an oversight in writing the code for the nested cross-validation and the way Keras handles layers, the exact combination of hyperparameters was not saved.

\begin{table}[H]
\centering
\begin{tabular}{rrrrr}
  \toprule
XY & Outer Loop & Hyperparameter Set & Inner Loss & Outer Loss \\ 
  \midrule
  Off & 1 &  48 & 2889.41 & 3558.67 \\ 
    Off & 2 &  89 & 3642.95 & 5917.40 \\ 
    Off & 3 &  48 & 3151.67 & 6036.90 \\ 
    Off & 4 &  48 & 2798.50 & 13492.27 \\ 
    Off & 5 &  48 & 2848.46 & 7047.40 \\ 
    \midrule
    Off & & Mean & 3066.20 & 7210.53 \\ 
  \hline
  \hline
  On & 1 & 138 & 3193.40 & 3270.52 \\ 
   On & 2 &  22 & 3789.84 & 2756.68 \\ 
    On & 3 &  55 & 3569.40 & 5668.05 \\ 
    On & 4 &  55 & 3006.54 & 4940.63 \\ 
    On & 5 & 138 & 2954.60 & 4625.91 \\ 
   \midrule
   On & & Mean & 3302.76 & 4252.36 \\ 
   \bottomrule
\end{tabular}
\caption{\label{tab:ncv_outer} Outer Losses of the nested cross-validation, where inner loss is the loss of the best performing hyperparameter set per outer fold, and outer loss is the loss of that same hyperparameter set when evaluated against the outer test set}
\end{table}

It is fair to compare these values with those from the repeated cross-validation for disaggregation (Table~\ref{tab:cv}), as in both cases it was a 5-fold cross validation. The loss for disaggregation with the field and IID turned off was 3843.41, and the loss for disaggregation with the field and IID turned on was 2852.13. Despite the hyperparameter tuning of the Kedis model, disaggregation still performed better for out-of-sample loss in both cases.

\section{Execution Time Comparison}
\label{sec:tt_res}

\begin{figure}
    \centering
    \includegraphics[width=\textwidth]{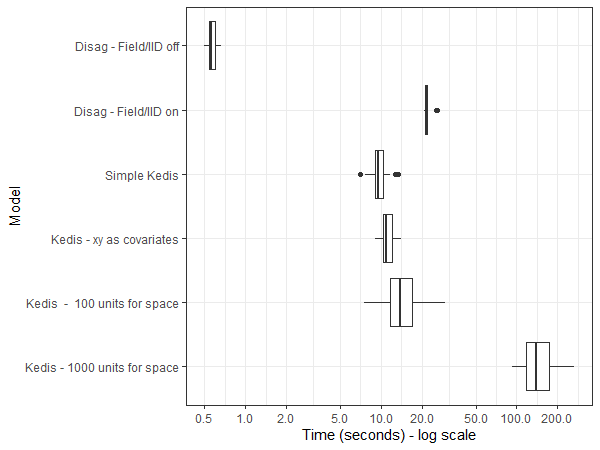}
    \caption{Boxplot of the execution times for each tested model}
    \label{fig:tt_box_log}
\end{figure}

Table~\ref{tab:exectime} outlines the summary statistics for the execution time measurements for all six models. Figure~\ref{fig:tt_box_log} displays the distribution of the execution times on a boxplot.
 
\begin{table}[H]
\centering
\begin{tabular}{lccccc}
  \toprule
Model & n & mean & sd & \multicolumn{2}{c}{95\% CI} \\ 
  \midrule
Disaggregation - Field/IID off &  50 & 0.57 & 0.04 & 0.56 & 0.58 \\ 
  Disaggregation - Field/IID on &  50 & 22.24 & 1.65 & 21.77 & 22.71 \\ 
  \addlinespace
  
    Simple Kedis &  50 & 9.69 & 1.25 & 9.34 & 10.05 \\ 
      Kedis - xy as covariates &  50 & 11.22 & 1.22 & 10.87 & 11.56 \\   Kedis  -  100 units for space &  50 & 14.96 & 4.91 & 13.57 & 16.36 \\
  Kedis - 1000 units for space &  50 & 148.12 & 40.90 & 136.49 & 159.74 \\ 

   \bottomrule
\end{tabular}
\caption{\label{tab:exectime} Summary statistics of the execution time for each model. Times are in seconds.}
\end{table}

The disaggregation model with no field of IID is by far the fastest, with a mean execution time of 0.57s. The slowest model is the Kedis model with 1000 units in a hidden layer for coordinate training.

\section{Uncertainty Estimates with Monte Carlo Dropout}
\label{sec:uncertain_res}

\begin{figure}
    \centering
    \includegraphics[width=\textwidth]{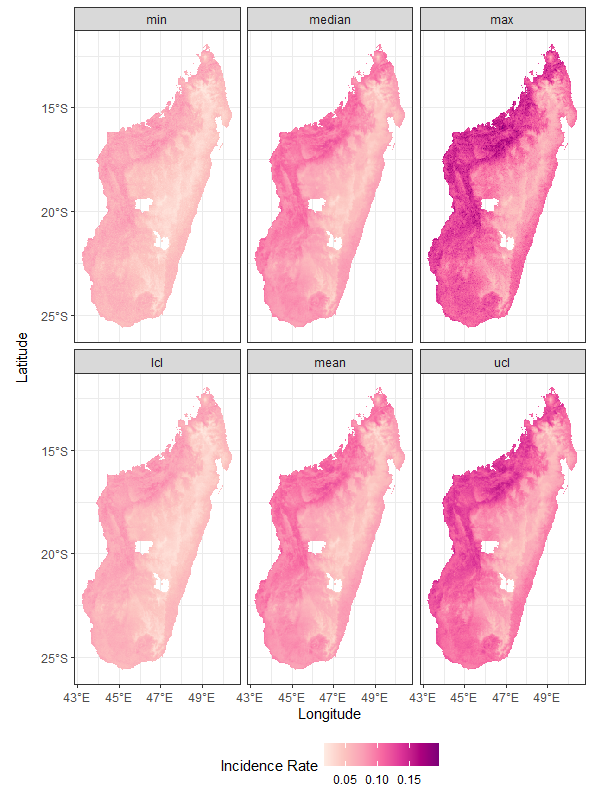}
    \caption{Model metrics from an example Monte Carlo dropout simulation}
    \label{fig:uncertainty}
\end{figure}

To assess the feasibility of Monte Carlo (MC) dropout, a simple design study was conducted. A simple Kedis model was built and sampled 100 times. Each pixel within the stack of 100 was assessed for mean, median, min, max, and standard deviation. From these values, Figure~\ref{fig:uncertainty} could be built which displays the uncertainty metrics of the test. Confidence limits were calculated as 1.96 standard deviations either side of the mean.

\begin{figure}
    \centering
    \includegraphics[width=\textwidth]{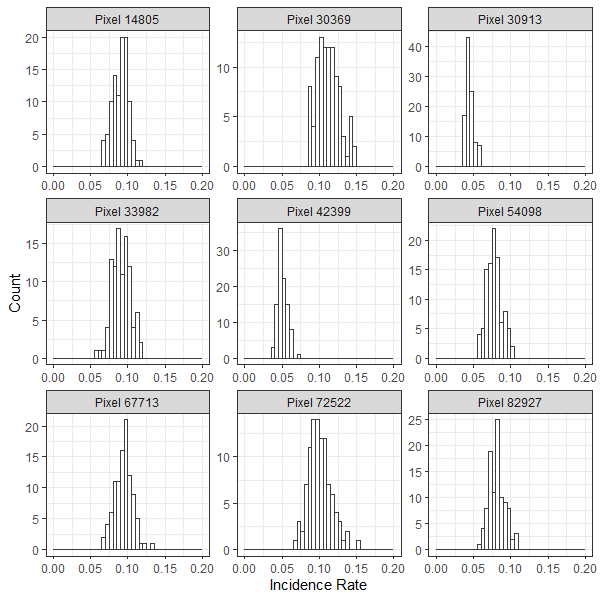}
    \caption{Histogram of nine randomly selected pixels from the MC dropout simulation}
    \label{fig:mc_hist}
\end{figure}

\begin{figure}
    \centering
    \includegraphics[width=\textwidth]{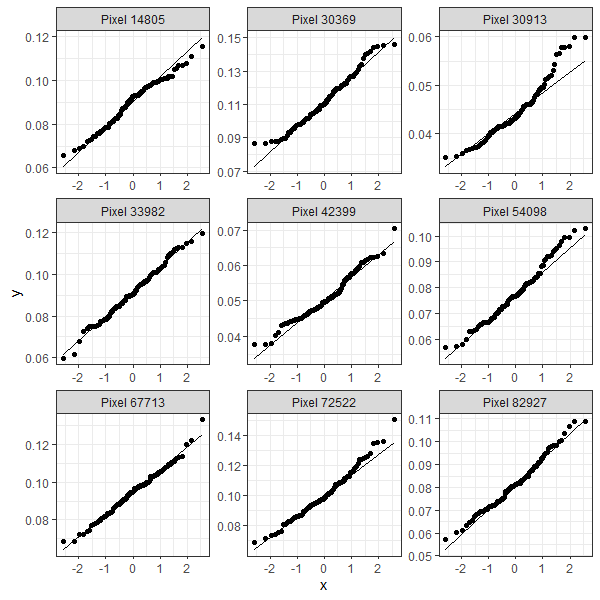}
    \caption{Q-Q plot of nine randomly selected pixels from the MC dropout simulation}
    \label{fig:mc_qq}
\end{figure}

The distribution of nine randomly selected pixels can be viewed in Figure~\ref{fig:mc_hist}, and the corresponding Q-Q plot is Figure~\ref{fig:mc_qq}. These plots show the distribution of each pixel is well described by a normal distribution.

\chapter{Discussions}
\label{ch:discussions}
This project started with a hypothesis that a neural network could be used to perform disaggregation regression; the primary motivation behind this was that neural networks may offer a solution to the problems of poor accuracy, no flexibility and slow speed that surround disaggregation currently. The novel method, 'Kedis' (Keras-Disaggregation) was successfully developed and benchmarked against traditional disaggregation on the metrics used as motivation for the project. Kedis was written entirely in R, and a package has been developed and deployed on GitHub that allows it to be used in production.

\section{Accuracy of Kedis}

The neural network implementation of disaggregation replicates GLMs through the inclusion of a linear predictor and link function. If no hidden layers are specified, it should replicate the behaviour of disaggregation with spatial effects turned off. This was directly compared in Section~\ref{sec:direct_comp_res}, and showed that although Kedis did make sensible predictions, there were some differences with disaggregation. As the Kedis model had no hidden layers, only five weights were available for training which directly corresponded to the coefficients of a linear disaggregation regression. These weights were different in each case, although the models created a visually similar pattern of disaggregation. The accuracy of the model was determined by reaggregating the output with the population as an offset over the borders of each region, then comparing to the reported incidence rate used to train the model. These incidence rate comparisons are displayed as a scatter plot in Figure~\ref{fig:direct_scatter}, and show that Kedis struggles to make accurate predictions when the true incidence rate is low. Both Kedis and disaggregation underestimate regions with very high rates. This is a limitation of both methods, as in disaggregation is it important to be able to pinpoint areas with high incidence, if the purpose is to optimise resources towards those regions. Unexplained variance is again suggested in Table~\ref{tab:direct_summary}, where the reported standard deviation for disaggregation and Kedis is half and a quarter of the actual standard deviation, respectively. This suggests that the covariates are insufficient to accurately describe the observed incidence, although disaggregation still models the variance better than Kedis. Figure~\ref{fig:cor_disag_kedis} shows a good correlation between the disaggregation and Kedis models (0.71), although it is also clear from this plot that Kedis overestimates lower values and the agreement is better in higher values. A Bland-Altman plot could have been used to better assess the agreement between the two tests.

The repeated cross-validation assessed the same simple models, plus additional models with varying degrees of complexity for space. Disaggregation objectively performed better than Kedis in every case, with the worst disaggregation model having a mean Poisson loss of 1198 lower than the best performing Kedis model. This agrees with the in-sample comparisons made between the simple models previously, although the loss is quantified in this case. An unexpected result is that adding more nodes to the handling of space in Kedis was detrimental to the model. This is opposite to the effect in disaggregation, where with the field turned on the loss was 991 lower. This effect may be explained by the normalisation of coordinates in Kedis before model fitting. In Kedis, the coordinates are scaled and centred in the \verb|prepare_data| function to their own mean and standard deviation. This is a limitation of how Kedis has been implemented, although it should be simple to resolve by allowing custom normalisation parameters in the data preparation function. In cross-validation, each fold would have a different mean and standard deviation for latitude and longitude. With highly complex spatial relationships (such as one with 10,000 nodes), these differences could lead to unpredictable local effects across each fold and may explain the detrimental effect of wider layers.

Cross-validation is traditionally used to measure the ability of a model to generalise well to new data. Disaggregation is not typically used to make predictions over areas where a response variable is unknown. It seems therefore that a model that fits well in-sample will make powerful predictions regardless of the out-of-sample accuracy. Indeed, a Kedis model with a highly complex set of hidden layers could learn a relationship which describes the data very well in-sample, and would therefore have a low loss. The issue arises when visualising the disaggregated output from this model: the complex relationships would lead to a small number of pixels with an extremely high incidence accounting for almost all of the cases and the rest being close to zero. This is obviously incorrect. A cross-validation of this model would confirm a high out-of-sample loss and is a better indicator of the model accuracy. \citet{arambepola2021} investigated the power of cross-validation on predictive performance with a simulation study and found that "cross-validation correlation on the aggregate level was a moderately good predictor of fine-scale predictive performance".

\section{Flexibility}

A major benefit of Kedis over disaggregation is the improved flexibility for a user to specify complex models. Nested cross-validation was used to assess the out-of-sample accuracy of an optimal set of hyperparameters. A set of hyperparameters was generated by a grid search for one layer and a random search for two to four layers. The hyperparameter set was tested with two nested cross-validation models, one with coordinates as covariates and one without. The best hyperparameter set for each inner loop was refitted on the respective outer training set, and validated against the test set. Assessing just the inner loops, the loss is a vast improvement over the simple model cross-validation discussed above, although this is not the primary outcome of a nested cross-validation. The model with latitude and longitude as covariates had a greater loss than the one without, again signalling there is an issue with the handling of spatial data that needs addressing. The loss of the outer loop for the model without coordinates was 7210 and the model with coordinates was 4252. The increased loss on the  outer loop is expected, as the best hyperparameter set on the inner loop is not necessarily the best set for the outer loop. Even after the hyperparameter tuning, the Kedis models still perform worse than the disaggregation models that were cross-validated in Section~\ref{sec:cv_res}. The decision to limit the number and type of hyperparameters to investigate to covariate training layers was made consciously due to the extremely high computational burden and time required to perform a nested cross-validation. Other hyperparameters that could have been investigated include different layer types and orders, extending the width of the dense layers, including layers for the \verb|layers_xy| option and different optimisers (including parameters within optimisers such as learning rate). An interesting outcome of the nested cross-validation could be to visualise any relationships between changing parameters and loss, e.g. how the loss is affected by dropout. Although this would have been possible with the models that were fitted, the behaviour of the Keras R package is that is does not easily allow models to be used over multiple sessions. In practice, this meant that the actual hyperparameter sets that were used were lost once the R session was closed.

\section{Execution Time}

\begin{figure}
    \centering
    \includegraphics[width=\textwidth]{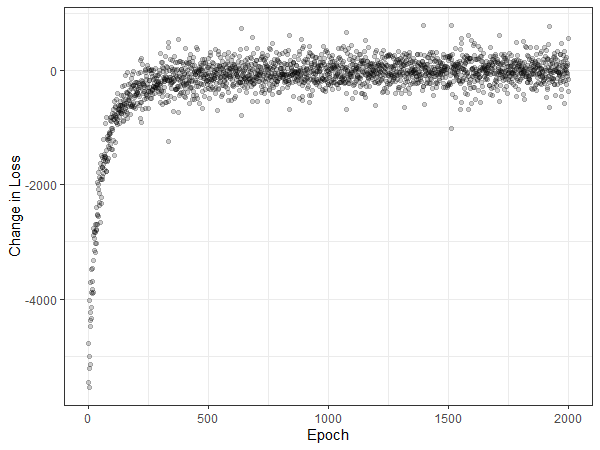}
    \caption{The change in loss by epoch of an example Kedis model trained over 2,000 epochs}
    \label{fig:loss_change}
\end{figure}

The execution time of the Kedis models varied depending on the number of layer, epochs, stopping criteria, optimiser, and many other factors. The Kedis model used for direct comparisons in Section~\ref{sec:tt_res} used an Adam optimiser \citep{kingma2014} and the default option for any hyperparameters. The Kedis models (with the exception of the one with 1000 units for training covariates) all performed at an acceptable speed, and were slightly faster than the disaggregation model with field and IID on. The fastest model was disaggregation without field or IID, which was almost instant. The choice of early stopping threshold also massively affects the execution time of a neural network. Figure~\ref{fig:loss_change} shows the change in loss per epoch of an example Kedis model and demonstrates the idea of diminishing returns. Choosing a sensible stopping criteria will ensure that the benefit of neural networks are realised without wasting time by training over unnecessary epochs. The early stopping callbacks all used arbitrary values in this study, and future work could investigate methods of tuning these parameters and could improve execution time by a substantial magnitude.

A major speed benefit of using Keras that was not looked into was the native support for hardware acceleration and parallel execution. GPUs are commonly used in deep learning applications due to the much higher core count than CPUs and can efficiently run in parallel. TensorFlow (and as a result Keras) supports using CUDA enabled GPUs from NVIDIA \citep{tensorflow_cuda} so realising the benefits of hardware acceleration is relatively simple, despite not investigated in this study. The disaggregation package currently does not support any method of hardware acceleration and only offers parallel computing with functionality provided by the 'parallel' \citep{parallel} and 'foreach' \citep{foreach} packages.

\section{Uncertainty Modelling}
Uncertainty modelling with Kedis was briefly investigated using a Monte Carlo dropout method. This method has the advantage that is computationally efficient: it trains one model but predicts from that model a number of times. It is also simple to implement, as Keras natively provides functionality to "trick" dropout layers into believing inference is actually forward propagation \citep{oleszak_2021}. In the example provided in Section~\ref{sec:uncertain_res}, a simple Kedis model with one dropout layer was defined and sampled from 100 times. The plots of predictions appear plausible and the nine random pixels assessed appear to be normally distributed (Figure~\ref{fig:mc_qq}). Further work is required to study the behaviour, accuracy and stability of the MC dropout method in terms of disaggregation modelling and to investigate any potential limitations of this method. 

MC dropout relies on models having dropout layers. Although this is not uncommon, it reduces some flexibility that is a primary benefit of the Kedis method. It is not immediately obvious how the dropout rate affects the range of uncertainty (does a higher dropout lead to a wider prediction interval?), or what effect stacking multiple dropout layers has on the uncertainty prediction. The current implementation uses the same dropout layers for training and inference; it may be possible to stack two independent dropout layers and use one for training and one for inference, or to include a dropout layer for inference only in a model that doesn't benefit from the improved generalisation during training. The intention of a method like this would be to optimise the dropout for training and inference separately and stabilise the predictions. This behaviour is not natively supported by Keras, although it should be possible to implement.

Keras has a set of regularisation layers that are only active during training time that can be exploited in some way to create model uncertainty. More in depth research would be required to truly understand the power of these layers. Gaussian noise and Gaussian dropout layers are the two layers with most potential to aid uncertainty predictions. Gaussian noise generates random values from a Gaussian distribution with a standard deviation set by the user, and adds that value to the weight. In Kedis modelling, this would likely be adding a random noise of pre-specified standard deviation to the linear predictor portion of the function. For uncertainty predictions, Gaussian noise may be unsuitable as the uncertainty predictions may simply have a standard deviation that is equal to the standard deviation of the additive noise. Gaussian dropout also draws samples from a Gaussian distribution with a pre-specified standard deviation, but multiplies the weights by that value instead of performing addition. Gaussian dropout should not suffer this same problem as Gaussian noise it is multiplicative, although this needs further study. In addition to the Keras in-built regularisation layers, other types of dropout exist as described by \citet{thevenot_2022}: DropConnect \citep{dropconnect}, which applies dropout to the weights and biases as opposed to the nodes directly, Standout \citep{standout}, where the probability of dropout is adaptive according to the weights, or a number of other dropout techniques which are not applicable in this case. Each could have applications in uncertainty modelling although the exact behaviour is not immediately clear.

An alternative method of estimating uncertainty in a neural network is deep ensembles. Deep ensembles were introduced by \citet{deepensemble} as a method of estimating uncertainty that is "simple to implement, readily parallelizable, requires very little hyperparameter tuning, and yields high quality predictive uncertainty estimates". The method involves training a number of models (an 'ensemble') and using the spread of the predictions as a score of uncertainty. Bootstrapped aggregation (or 'bagging') can be used to train each model with a subset of the original set, although this method may suffer problems with the loss of data in small sample sizes (such as in disaggregation, where the sample size is the number of regions). \citet{lee2015} suggested that training an ensemble on the entire dataset with different random initialisers is superior to bagging, although that study was interested in assessing predictive accuracy as opposed to uncertainty. Random forests use a subset of the features in training to improve generalisability, although this suffers the same issues as bagging due to the data loss of a small initial sample. The method examined by \citet{deepensemble} involves using the whole dataset, and minimising the loss using an adverserial example \citep{szegedy2013} for each model in the ensemble. Using adverserial examples, as opposed to the whole dataset, to compute the loss is more efficient, as the focus is on regions where the loss (and therefore gradient) is high. The \citet{deepensemble} paper compares deep ensembles using this method to MC dropout \citep{gal2016} and probabilistic back-propagation, and showed that deep ensembles predicted similar levels of uncertainty to MC dropout. A limitation of this method compared to MC dropout is that it still involves training multiple models. \citet{deepensemble} does not mention a study of computation time: an important motivation for using neural networks to replace disaggregation, but it is likely that it would be slower than MC dropout.

\section{Alternative Methods of Handling Spatial Autocorrelation}

The disaggregation package uses Integrated Nested Laplace Approximation with Stochastic Partial Differential Equation (INLA-SPDE) to approximate the Gaussian process in order to describe the correlation between geographically close areas. Accounting for spatial autocorrelation "is a both a nuisance (sic), as it complicates statistical tests, and a feature, as it allows for spatial interpolation" \citep{hijmans2021}. In Kedis, coordinates may be used as covariates, or alternatively trained on their own separate layers and added to the linear predictor, similar to how the disaggregation package handles the Gaussian process. "It has long been known that a single-layer fully-connected neural network with an i.i.d. prior over its parameters is equivalent to a Gaussian process (GP), in the limit of infinite network width" \citep{lee2017}. With this in mind, it is possible to approximate a Gaussian process in Kedis by including a densely connected layer with a very high number of nodes for the \verb|layers_xy| term in \verb|build_model| function:

\begin{verbatim} 
layers_xy <- list(layer_dense(units = 100000, 
                              activation = "relu"))
\end{verbatim}

In practice, using a very wide layer in Keras is not trivial, as there are limits to system memory and it is very slow. As shown in the Execution Time Comparisons section of Chapter~\ref{ch:results}, increasing the number of nodes in a layer increases training time in a linear fashion (\(\mathcal{O}(n)\), where \(n\) is the number of nodes in a single hidden layer).

Care should be taken when making predictions in a model that includes space. As already discussed, the normalisation coefficients that scaled the spatial data in training should also be used when preparing data for predictions. Additionally, it may not be sensible to extrapolate beyond the coordinates used in training. A Gaussian process is a complex function that describes the continuous spatial field by using a discrete set of points, so making predictions outside the range of the training set would lead to unpredictable behaviour.

Designing alternative methods for dealing with spatial autocorrelation is a whole field of research, and is beyond the scope of this project. The option to separately train layers for coordinates allows great flexibility to implement specific solutions, and is a strength of Kedis.

\section{Model Assumptions and General Problems with Disaggregation}

A traditional Poisson regression for modelling disease uses the assumptions of:

\begin{itemize}
    \item The response is a count variable with a constant exposure, or the exposure can be measured and is known. Exposure can be time, space or population.
    \item The mean is equal to the variance at all covariate levels.
    \item  Every disease case is independent from every other case. That is, the occurrence of a disease case has no effect on the likelihood of any other case.
\end{itemize}

Although disaggregation is a new concept, the assumptions for the underlying GLM should still hold. In disaggregation, all of these assumptions can be shown to some extent to be not true, or difficult to quantify.

The assumption of modelling rates based on cases and population is an extension of a Poisson regression. For this to hold true, each region should have a consistent population throughout the period observed (one year). This is not likely to be true, as populations fluctuate due to births, deaths and migration. The population reported for each region is a snapshot at a particular time so is used as a proxy for a true exposure variable (for example person years) which would be much more difficult to quantify.

The assumption that the mean is equal to the variance is central to a traditional Poisson regression, and methods exist to deal with cases where the variance is (usually) much larger than the mean. This occurs usually in poorly specified models, where there is variance that is unexplained by the available covariates. Quasi-Poisson and Negative Binomial models offer a solution to this overdispersion by allowing the variance to grow as a function of the mean, although implementing these in a Kedis model may be non-trivial. The disaggregation package deals with the overdispersion by fitting a multi-level model with an IID random effect at the observation level; a Kedis implementation of this was not investigated and is cause for future research.

Another explanation for the overdispersion in modelling diseases is the reality that cases are not independent. Malaria is a communicable disease, therefore the presence of a high number of cases locally may increase the chance of becoming infected; although there is no direct human-to-human transmission, the chance of mosquitoes biting an infected person then going on to infect a further person is higher in areas with high incidence. This is partially explained by the spatial autocorrelation discussed earlier, and is a limitation to the current implementation of Kedis. Likewise, a random effects model may go some way to explain the inherent clustering of cases. The effect of migration of infected persons is also ignored, and this is a major contributor to the spread of disease \citep{gushulak2004}, as clearly observed in the SARS-CoV-2 pandemic \citep{zargari2022}.

Both the disaggregation package and Kedis focus on spatial disaggregation. Another type of disaggregation is 'temporal disaggregation', which models incidence as a function of time. Malaria cases are aggregated over space, but also have been aggregated over a whole year. As covariates are available both at high spatial resolution but also a high temporal resolution, it may be possible to disaggregate in three dimensions and include lag effects. The disaggregation models presented in this project represent the average incidence over a year instead of an immediate risk score.

\section{The 'Kedis' R Package}

The Kedis R package was developed as part of this project and allows the techniques described here to be easily implemented in practice. Although the core methods in the package work properly and are stable, there are some minor functions (such as those for performing Cross-Validation) that require some work to be stable. In addition, writing proper documentation was not a priority during the early stages so it is fairly sparse. Unit tests have also not been written, which are an essential part of package development to ensure stability and consistency across patches and updates. The addition of an option to force the model to work in training mode would also be useful to implement MC dropout, as would wrapper functions to perform MC sampling and processing of the results (although this is fairly straightforward to implement manually). 

\section{Summary}

This project implemented disaggregation modelling with a neural network, and tested it against the disaggregation R package on metrics of accuracy, flexibility and execution speed. The new method, Kedis, did not offer the predictive performance of disaggregation, but was a substantial improvement in flexibility and speed. Kedis performed poorly at predicting areas with extremes of incidence rate, but predictions correlated well with previous methods and has more flexibility for extensions that may improve model accuracy in the future.

\backmatter

\bibliography{bib/bibliography, bib/r_packages}

\end{document}